\documentclass[aps,prc,groupedaddress,twocolumn,showpacs,longbibliography,nofootinbib]{revtex4-1}

\usepackage{amsmath,amssymb,bm}
\usepackage{ae}

\newcommand{\frc}[2]{\mbox{$\frac{#1}{#2}$}}

\newcommand{\pp}[1]{%
}

\usepackage{graphicx} 

\DeclareMathSymbol{\NS}{\mathord}{AMSb}{"4E}

\newcommand{\expect}[1]{\ensuremath{\langle{#1}\rangle}}

\newcommand{\op}[1]{\ensuremath{#1}}

\renewcommand{\vec}[1]{\ensuremath{\bm{#1}}}

\newcommand{\rO}{\ensuremath{\op{r}}}

\newcommand{\QO}{\ensuremath{\op{Q}}}

\begin{document}


\title{Isoscalar and neutron modes in the $E1$ spectra of Ni isotopes and the relevance of shell effects and the continuum}

\author{P.~Papakonstantinou} 
\email[]{ppapakon@ibs.re.kr}
\affiliation{Rare Isotope Science Project, Institute for Basic Science, Daejeon 305-811, Korea}

\author{H.~Hergert}
\affiliation{National Superconducting Cyclotron Laboratory, Michigan State University, East Lansing, MI 48824-1321, USA} 

\author{R.~Roth}
\affiliation{Institut f\"ur Kernphysik, Technische Universit\"at Darmstadt, 64283 Darmstadt, Germany}

\date{\today}

\begin{abstract}
%
%
%
We study theoretically the electric dipole transitions of even Ni isotopes at low energies, using  
the self-consistent quasi-particle random-phase approximation with the D1S Gogny interaction and a continuum-RPA model with the SLy4 Skyrme force. 
We analyze isoscalar states, isovector states, and the dipole polarizability. 
We define a reference value for the polarizability, to remove a trivial dependence on the mass number. 
We compare our results with data and other calculations, with a focus on collective states, shell effects, and threshold transitions.  

Our results support the presence of a strong isoscalar transition, with little or moderate $E1$ strength, as a universal feature of ordinary nuclei. 
In moderately neutron-rich Ni isotopes, namely $^{68}$Ni and neighboring isotopes, this transition is found bimodal due to couplings with surface neutrons. 
An adequate treatment of the continuum states appears essential for describing suprathreshold $E1$ strength, especially beyond $^{68}$Ni. 
Very exotic isotopes ($N>50$) are found highly polarizable, with practically all their $E1$ strength in the continuum. 
The dipole polarizability and the neutron-skin thickness are influenced by shell structure in different ways, so that they can appear anticorrelated. 
A comparison with existing results for lighter (Ca) and heavier (Sn) nuclei suggests that 
the so-called pygmy dipole strength is influenced strongly by shell effects and that, partly for that reason,  
its isospin structure depends on the mass region. 
%
\end{abstract}


\maketitle



\section{Introduction}

The electric dipole transitions of nuclei at low energies, below the giant dipole resonance, have been under intensifying investigation for more than two decades~\cite{SAZ2013,PVK2007,KrS2009}. 
Knowing or predicting correctly the concentration of dipole strength, which cannot be attributed to the low-energy tail of the giant dipole resonance, is especially consequential in describing nucleosynthesis processes~\cite{AGT2007}. 
Through its correlations with other observables, the amount of strength may be used to constrain the isovector properties of the nuclear equation of state~\cite{ReN2010,Pie2011}. 

In recent years, the use of hadronic probes and the investigation of unstable nuclei have offered new insights into the nature and isospin structure of low-energy dipole transitions~\cite{SAZ2013}.   
The extraction of complete spectra on both sides of the particle emission threshold also constitutes important progress~\cite{Tam2011,Has2015X}. 
The effort is bound to continue, both theoretically and experimentally. 
New results on exotic nuclei are under analysis and facilities with improved capabilities are planned. 
Moreover, the already available dipole spectra are so complex, as functions of the excitation energy, mass, and isospin asymmetry, 
that a unified theoretical interpretation of them has been elusive. 
New data will be needed to resolve the situation, as well as the use of diverse nuclear models.  

Various factors and mechanisms have been proposed or are recognised as relevant for generating low-energy dipole transitions in medium-mass and heavy nuclei. 
A variety of microscopic models have been employed in related studies, mainly the quasi-particle or plain random-phase approximation (QRPA or RPA) and models considering phonon coupling. 
The presence of strength has been attributed to 
the concentration of single-particle transitions in the $1\hbar\omega$ energy region~\cite{OHD1998,KrS2009,Roc2012,ReN2013,Bar2013}, 
an alpha-cluster mode~\cite{IaJ1982,*Iac1985,SPZ2015},  
a surface-dipole vibration, akin to the other known quadrupole and octupole vibrations~\cite{Pap2015,PPR2011,PHP2012,Der2014,PHP2014} and of toroidal structure~\cite{Urb2012}, 
or, in the case of neutron(proton)-rich nuclei, an exotic neutron(proton)-skin oscillation~\cite{PVK2007}. 
The fragmentation patterns and precise strength can be explained by the coupling of the dipole transitions to collective phonons~\cite{LRT2013,Pon2014,Kna2014,Ach2015} 
or collisional damping~\cite{GGC2011}.  
Is has been demonstrated, in studies along isotopic chains, that also shell structure and deformation influence the low-energy dipole strength~\cite{ENI2014,Mas2014,PHP2012,PHP2014}. 
We may add that threshold particle resonances are a general feature of potential barrier scattering and may also contribute to the photoneutron cross section 
just above the emission threshold.  
The question now is, whether certain mechanisms dominate in the different mass and asymmetry domains of the nuclear chart, and at what energy in relation to the emission threshold. 

The answer to the first part of the question may well be different in the case of the isoscalar (IS) or isovector (IV) channel. 
That is the conclusion we have reached in Refs.~\cite{PPR2011,PHP2012,Der2014,PHP2014,Pap2015} for the dipole strength below particle-emission threshold 
and in moderately neutron-rich nuclei 
(below critical shell closures). 
The IS strength is attributed to a collective mode of vibration, which is excited in $N=Z$ nuclei too; the IV strength, largely to single-particle transitions. 
Our conclusions were based on QRPA calculations with the Gogny D1S interaction and comparisons with experimental data as well as with other calculations, 
along the Ca and Sn isotopic chains and in $N=Z$ nuclei.  
A genuine neutron-skin oscillation may be excited either above threshold or in extremely neutron-rich nuclei. 
We note that relativistic models favor a neutron-skin oscillation in moderately neutron-rich nuclei~\cite{VNP2012}, while 
IS coherence is generally the outcome of non-relativistic studies~\cite{Roc2012}. 
Such qualitative disagreements can be attributed to the softness of the symmetry energy, which is typically predicted to be different by the two types of models.  

Pinpointing the mechanisms which control the dipole strength and establishing correlations with other observables 
often requires extreme care in interpreting theoretical results, and of course data. 
A case in point is the correlation between the dipole polarizability and the neutron-skin thickness, which was found model-dependent~\cite{Pie2012}. 
One question which we will address in this work is whether the shell structure affects the polarizability and the skin thickness in different ways. 
As regards data, it will be very interesting to confirm experimentally the dramatic role of shell closure revealed in several theoretical studies. 

The second part of the question we have posed above is related to scattering in the particle continuum. 
In most theoretical studies so far the particle continuum is at best discretized (for recent exceptions see, e.g., \cite{DaG2012,Mat2015}). 
Broad resonances in the vicinity of the particle threshold may not be adequately described. 
This is another issue we will address here.

In the present work we focus on the Ni isotopic chain, which  
represents an intermediate mass region, between the lighter Ca and heavier Sn nuclei. 
Known isotopes span the whole $pfg_{9/2}$ neutron shell.  
They include the long-lived, symmetric $^{56}$Ni isotope and several stable ones. 
Photoresponse data exist for the latter, below particle threshold, and for the neutron-rich $^{68}$Ni, above particle threshold, while more are expected to be analysed soon. 
We are interested in possible collective excitations, 
the role of shell structure, and 
in general in the factors determining the amount of strength. 
As in previous work, we calculate the properties of dipole excitations using the QRPA and the Gogny D1S interaction. 
Furthermore, we use a continuum-RPA model (CRPA) and the Skyrme functional SLy4, to examine the threshold strength. 
We will compare our results with available data and other theoretical results. 

Our dipole-mode nomenclature should be clarified at this point. 
In the following we refer repeatedly to two types of collective dipole excitations. 
One is an ordinary {\em surface dipole vibration}, which corresponds to the isoscalar low-energy dipole (IS-LED) mode studied in Refs.~\cite{PPR2011,PHP2012,PHP2014,Der2014,Pap2015}. 
We will therefore denote it as {\em IS-LED}. 
The other is driven by neutron excess and we will refer to it as a {\em neutron mode}.  
The term {\em neutron-skin} mode or oscillation we reserve explicitly for the oscillation of the neutron skin as a spatially defined entity. 
Some authors call that the pygmy resonance. 
In this work, however, we use the term {\em pygmy} to refer to low-energy $E1$ transitions of any physical origin, as long as they are not considered fragments of the giant dipole resonance. 
 
The plan of this paper is as follows. 
In Sec.~\ref{Sec:Theory} we introduce briefly the theoretical models we use and define the quantities of interest. 
In Sec.~\ref{Sec:Results} we present and analyse our results. 
In particular, in Sec.~\ref{Sec:ResultsMain} we examine the isovector and isoscalar response of Ni isotopes and identify interesting low-energy structures to be studied next. 
In Sec.~\ref{Sec:ResultsSystematic} we examine the evolution of those structures with respect to the neutron number.  
Having identified two potentially interesting modes, in Sec.~\ref{Sec:Two} we analyze them through their transition densities and wavefunctions. 
In Sec.~\ref{Sec:Sums} we calculate the total low-energy strength and the dipole polarizability and compare our results with existing data. 
A critical discussion of our findings is presented in Sec.~\ref{Sec:Discuss}. 
We focus on the relevance of shell structure and the phenomenon of isospin splitting and  
we comment on how our results fit into the ongoing discourse about the relation of dipole states with the softness of the symmetry energy.  
We summarize our conclusions in Sec.~\ref{Sec:Summary}.

\section{Theory background\label{Sec:Theory}} 

\subsection{Theoretical models} 

In this work we employ two models based on the random-phase approximation. 
The models have been discussed in detail elsewhere. 
Therefore, we will now present them only briefly and refer the reader to respective publications for more details. 

We use mainly the self-consistent quasi-particle random-phase approximation (QRPA) with the Gogny D1S interaction~\cite{BGG1991}, 
which has yielded already interesting results in our previous studies of dipole strength~\cite{PPR2011,PHP2012,PHP2014}. 
The QRPA is formulated in the canonical basis of the Hartree-Fock-Bogolyubov model (HFB) describing the ground state. 
The same Hamiltonian is used in the description of the ground states and in the $ph$ and $pp$ channels in QRPA, 
leading to an excellent degree of self-consistency. 
The theory and implementation are presented in detail in Ref.~\cite{HPR2011}. 
The reader may also consult Ref.~\cite{PHP2012} for a concise presentation. 
A harmonic oscillator basis of 15 shells is used in the present work. 
The oscillator length parameter is chosen for each isotope such that the energy of the HFB ground state is minimized. 
 
Our QRPA model in the harmonic-oscillator basis lacks a proper treatment of the continuum. 
For this reason we also explore the continuum random-phase approximation (CCRPA) for Skyrme interactions. 
We employ the SLy4 parameterization~\cite{Chabanat:1997qh}, because it 
predicts similar nuclear-matter properties of relevance as the D1S model, namely the symmetry energy and the nucleon effective mass. 
Both models are asy-soft. 
They differ mainly in the Thomas-Reiche-Kuhn (TRK) enhancement factor, 
which influences the energy of the GDR, but not strongly the pygmy-strength properties or the neutron skin~\cite{Rei1999,ReN2010,ReN2013}. 

The CRPA formalism makes use of Green's functions. 
The ground state is obtained with the Skyrme-Hartree-Fock model~\cite{ReXX}. 
The unperturbed $ph$ propagators and the residual interaction are evaluated in coordinate space. 
Convergence issues with respect to the model space are thus automatically absent.  
Appropriate boundary conditions account for the possibility of particle emission when the particle energy is positive. 
The escape effect can be turned off by imposing box boundary conditions, namely that all wave functions vanish at some distance from the nucleus. 
The model is the same as that used in Ref.~\cite{PPP2005} and the reader is referred there for further information. 
The CRPA does not include pairing and we will use it only for a few isotopes which afford a description as closed-shell configurations. 
We typically use a radial mesh of 18~fm and a step size of 0.08~fm. 

\subsection{Quantities of interest} 

The transition strength distribution corresponding to an external field $\QO$ is given, in QRPA, by 
\begin{equation} 
S(E) = \sum_{\nu > 0} B(E_{\nu}) \delta (E-E_{\nu}) 
, 
\end{equation}  
where the transition strength of the $\nu -$th eigenvalue 
\begin{equation} 
B(E_{\nu}) = \left| \sum_{q<q'} b_{qq'}^{\nu} \right|^2 
\label{Eq:BofEQRPA}  
\end{equation} 
is given in terms of the 2-quasiparticle $X,Y$ amplitudes and the occupation probability factors $u,v$, 
\begin{equation} 
b_{qq'}^{\nu} = (X_{qq'}^{\nu} - Y_{qq'}^{\nu})(v_qu_{q'}-v_{q'}u_q) \langle q | \QO | q' \rangle  
\label{Eq:bqqnu}  
. 
\end{equation} 
The minus sign follows from $J^{\pi}=1^-$.  
In CRPA the strength distribution is determined by the polarization diagram via the Green's function 
\begin{equation} 
S(E) = -\frac{1}{\pi} \Im 
\sum_{t,t'} 
\int d^3 r d^3 r' 
f_t(\vec{r}) 
G_{tt'}(\vec{r},\vec{r'};E) 
f_{t'}(\vec{r}') 
,
\end{equation}  
where $f_t(r)$ 
denotes the spatial and isospin dependence of the external field. 

The isoscalar (IS) and isovector (IV) electric dipole response is determined in the two models, in the long-wavelength limit, by the following respective operators, 
which are corrected for spurious center-of-mass effects: 
\begin{equation}\label{eq:E1_IS_mod}
  \QO_{\mathrm{IS}}=e\sum_{i=1}^A\left(\rO_i^3-\frac{5}{3}\expect{\rO^2}r_i\right)\sqrt{3}Y_{1M}(\hat{\vec{r}}_i)\,,
\end{equation}
and
\begin{equation}\label{eq:E1_IV_mod}
  \QO_{E1}=e\frac{N}{A}\sum_{p=1}^Zr_pY_{1M}(\hat{\vec{r}}_p)-e\frac{Z}{A}\sum_{n=1}^Nr_nY_{1M}(\hat{\vec{r}}_n) 
\, .\end{equation}
As far as intrinsic excitations are concerned, the operator $\QO_{E1}$ is equivalent to the uncorrected $E1$ operator and to the isovector dipole operator 
\[ 
 \QO_{E1} \Leftrightarrow e\sum_{p=1}^Z r_pY_{1M}(\hat{\vec{r}}_p) \Leftrightarrow \frc{e}{2}\sum_{i=1}^A \tau_3^{(i)} r_iY_{1M}(\hat{\vec{r}}_i)  
,
\]  
in an obvious notation. 
When we speak of {\em E1 strength} we will refer to the electric-dipole excitation strength $B(E1)\uparrow$ corresponding to the operator $\QO_{E1}$ 
and we shall use {\em IS strength} when we refer to the operator $\QO_{\mathrm{IS}}$. 

The QRPA transition strengths obtained with the corrected operators (\ref{eq:E1_IS_mod}) and (\ref{eq:E1_IV_mod}) 
are the same as with the uncorrected forms, except of course for the spurious state. 
The spurious state appears always very close to zero energy. 

Our CRPA implementation is not fully self-consistent, in that spin-dependent operators and the Coulomb term are omitted from the residual interaction. 
Therefore, we always scale the residual interaction by a factor such that the spurious state appears close to zero energy. 
The scaling factor ranges from 1.06 for $^{68}$Ni, to 1.11 for $^{48}$Ni, to the large values of 1.15 for $^{78}$Ni and 1.22 for $^{56}$Ni. 
The use of corrected operators is imperative. 

The discrete energy spectra obtained with QRPA can be smoothed with a Lorenzian of a given width, for presentation purposes. 
In CRPA the spectra are continuous and include the escape width. 
In order to visualize very narrow or bound structures, an artificial width can be introduced by adding an imaginary constant to the particle-hole energy. 
In our presentation of results, when we say that our QRPA or CRPA results have been smoothed by means of a given width parameter, 
we will imply the width at half maximum. 

The dipole polarizability 
\begin{equation} 
a_{\mathrm{D}} = \frac{\hbar c}{2\pi^2} \int \frac{\sigma_{\mathrm{abs}}}{\omega^2} d\omega 
, \end{equation} 
which is proportional to the inverse-energy-weighted sum of $E1$ strength, has been found to correlate with the neutron-skin thickness and the density-dependence of the symmetry energy~\cite{ReN2010}. 
It has been measured in $^{208}$Pb~\cite{Tam2011}, $^{120}$Sn~\cite{Has2015X}, and in $^{68}$Ni~\cite{Ros2013}. 
We will calculate this quantity along the Ni chain. 
In order to eliminate a trivial dependence on the mass number, 
we define a reference, collective value $a_D^{\mathrm{coll}}$, as the polarizability corresponding to the giant dipole resonance if it was a single peak 
with the empirical energy~\cite{Hav2001} 
\[ E_{\mathrm{GDR}} = (31.2A^{-1/3} + 20.6 A^{-1/6}) \,\, \mathrm{MeV}\] 
and exhausting $100\%$ of the classical TRK sum rule. 
This gives us
\footnote{After submitting this manuscript we learnt that a similar reference value has been defined by S.Ebata et al. for related studies~\cite{NaEPC}. 
}  
\begin{equation} 
a_D^{\mathrm{coll}} = \frac{\hbar c}{2\pi^2} \frac{60 NZ \mathrm{MeV mb}}{A E_{\mathrm{GDR}}^2} 
\label{Eq:aDcoll} 
\end{equation} 
neglecting the TRK enhancement factor. 
 

The transition density, between the ground state and an excited state, determines how the system couples to the external field at the given excitation energy and 
provides insight into the microscopic structure of the excited state. 
In QRPA, solved in a spherical basis, the radial part of the transition density is given by 
\[ 
\delta\rho_t^{\nu}  = \sum_{q<q',t_q=t'_q} (X_{qq'}^{\nu} - Y_{qq'}^{\nu})(v_qu_{q'}-v_{q'}u_q) R_q(r) R_{q'}(r) 
, 
\] 
where $R_q(r)$ is the radial wavefunction of the canonical state and $t_q$ refers to its isospin (proton or neutron). 
 
In our CRPA model, transition densities are evaluated as follows. 
In the vicinity of a resonance, at energy $E$, the Green's function is proportional to a product of transition densities~\cite{BeT1974}. 
In particular,  
\begin{equation} 
\Im G_{t,t'}(r,r';E) \propto \delta \rho_t (r;E) \delta \rho_{t'}(r';E),  
\label{Eq:factorize} 
\end{equation} 
where we have used spherical symmetry to separate out the angular dependencies, and $t$ denotes a proton or a neutron. 
Folding the above quantity with the external field, we get a function  
\begin{equation} 
\delta \tilde\rho^{(f)}_t(r;E) = \sum_{t'}\int \Im G_{tt'}(r,r';E) f_{t'}(r') dr' 
, 
\end{equation} 
which is proportional to $\delta \rho_t (r;E)$, if Eq.~(\ref{Eq:factorize}) holds. 
We observe that if we fold this with the external field $f_t(r)$, we obtain $\pi$ times the transition strength (per energy unit). 
The above factorization is not always perfect, but in any case the above procedure helps us obtain a quantity, 
loosely called a transition density, which determines how the nuclear density couples to the external field at a given energy. 
It may depend on $f_{t}(r)$, therefore, we use the subscript $f$.    

\subsection{Assessing collectivity\label{Sec:Coll}} 

Different complementary ways can be used to assess the collectivity of a QRPA eigenstate. 
One of them is the amount of 2-quasiparticle configurations which contribute considerably to the norm or to the transition strength (summation in Eq.~(\ref{Eq:BofEQRPA})) of a given QRPA eigenvalue. 
In the case of dipole states, care must be exercised when using the latter criterion, as the results depend on the excitation operator used (i.e., whether it is corrected or not)~\cite{PHP2014}. 
Therefore, in addition, we will employ single-particle units, in order to assess the collectivity of a given state. 
For $E1$ transitions the usual single-particle unit must be corrected for the center-of-mass motion, by multiplying the usual estimate~\cite{RS80} by the effective-charge factor squared, namely $(N/A)^2$ for proton states and $(Z/A)^2$ for neutron states~\cite{Hav2001}. 
We choose to use the single-neutron unit 
\begin{equation} 
B_{\textrm{s.n.}}(E1)=0.06445A^{2/3}(Z/A)^2e^2\mathrm{fm}^2 
\label{Eq:Bsn} 
. 
\end{equation}  

\section{Results, analysis, and comparison with data\label{Sec:Results}} 

\subsection{Main features of response\label{Sec:ResultsMain}} 

First we compare the general features of the response, in particular giant resonances, between the two models and with some data. 
The only isotopes which can be described using closed subshells and, therefore, lend themselves to calculations with both models are those with $N=20,28,40,50$, 
all of which are unstable. 
Relevant data are available for stable isotopes and for $^{68}$Ni. 
We will, therefore, use $^{58,68}$Ni for a first comparison of our results with data. 

The nucleon separation energies for the $N=20,28,40,50$ isotopes are calculated within the D1S model to be $0.41,7.31,8.59,5.59$~MeV, respectively. 
For $N=28,40$ the values given in the AME2012 evaluation~\cite{AME2012} are $7.167,7.793$~MeV, respectively. 
 
\begin{figure*}
\includegraphics[angle=-90,width=\textwidth]{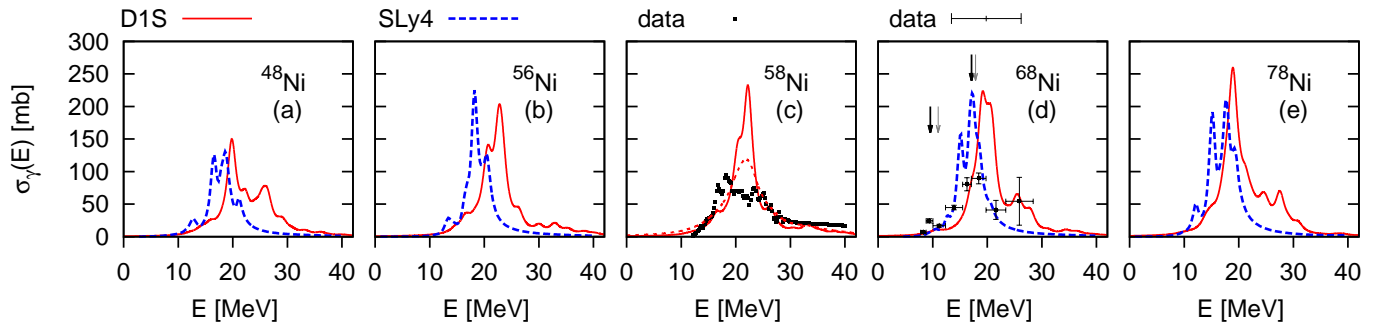}%
\caption{(Color online) Calculated photoabsorption cross section, for the indicated interactions and isotopes.  
The calculated cross sections are smoothed with a Lorenzian of width equal to 2~MeV. 
For $^{58}$Ni (c) the dashed curve corresponds to a smoothing width of 5~MeV. 
The data plotted in (c) and (d) for $^{58}$Ni and $^{68}$Ni are from Refs.~\cite{CDFE} and \cite{Ros2013,AuRPC}, respectively. 
The arrows in (d) indicate peaks identified in Refs.~\cite{Ros2013} (black) and \cite{Wie2009} (gray). 
\label{Fig:sigma}} 
\end{figure*}

In Fig.~\ref{Fig:sigma} we show the calculated photoabsorption cross sections for the above isotopes. 
A large smoothing factor has been used.  
The giant dipole resonance (GDR) is visible in all cases, with some degree of fragmentation. 
The GDR energies predicted by the two models are visibly different, by a few MeV. 
The difference can be attributed to the different TRK enhancement factor (0.6 for D1S, 0.25 for SLy4).  
An inspection of GDR data 
indicates 
that the QRPA+D1S model reproduces more correctly the higher-energy part and overall the mean energy of the GDR,  
while the CRPA+SLy4 model predicts rather correctly 
the main first peak of the GDR. 
We infer this from Fig.~\ref{Fig:sigma}(d)  
and from a juxtaposition of Figs.~\ref{Fig:sigma}(b,c) in the expectation that the GDR properties of $^{56}$Ni and $^{58}$Ni must be quite similar. 
Both models can be considered acceptable in this sense.  
Finally, they both appear to underestimate the fragmentation of the resonance. 
Indeed, the fragmentation and spreading of the GDR 
can be enhanced by effects beyond (Q)RPA, like collisional damping~\cite{Sch2010}, especially in heavy nuclei, and clustering~\cite{WBH2014}, especially in light nuclei. 
In medium-mass nuclei both mechanisms may contribute.  

\begin{figure*}
\includegraphics[angle=-90,width=\textwidth]{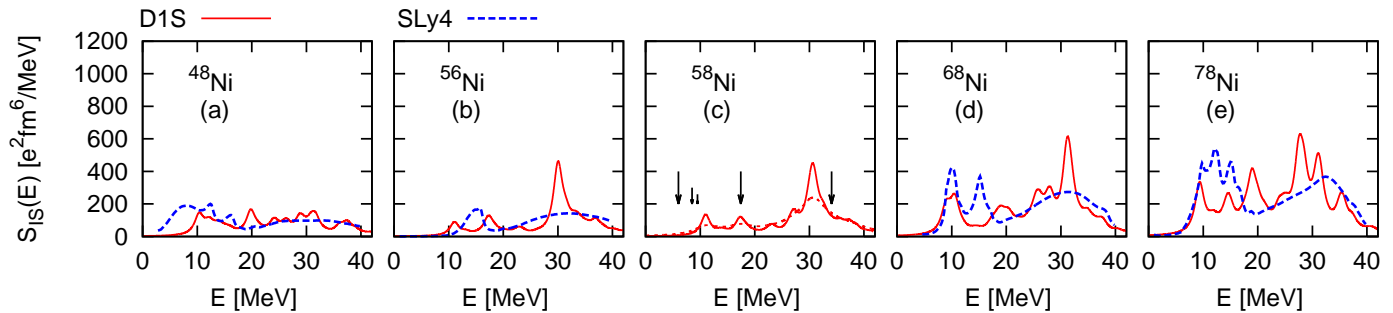}%
\caption{(Color online) Calculated isoscalar strength distribution, for the indicated interactions and isotopes.  
The distributions are smoothed with a Lorenzian of width equal to 2~MeV. 
For $^{58}$Ni (c) the dashed curve corresponds to a smoothing width of 5~MeV. 
The arrows show peaks detected below 10~MeV in $(\alpha ,\alpha'\gamma_0)$ experiments~\cite{Poe1992} 
and mean energies of gaussian fits to the IS strength distribution extracted from alpha scattering at small angles~\cite{Lui2006}. 
\label{Fig:ISGDR}} 
\end{figure*}

In Fig.~\ref{Fig:ISGDR} we show the calculated isoscalar strength distributions. 
In $^{58}$Ni IS dipole strength has been detected both below~\cite{Poe1992} and above~\cite{Lui2006} particle emission threshold 
and the major structures are indicated in Fig.~\ref{Fig:ISGDR}(c) with arrows. 

In all 5 isotopes, the broad structure at high energy, around 30~MeV, corresponds to the IS GDR, or dipole compression mode. 
We have confirmed that the proton and neutron transition densities show the typical behavior for this mode~\cite{Hav2001}, namely they are in phase and have a node. 
The smooth distribution given by the CRPA+SLy4 model in this energy region 
is not the result of a large smoothing factor, but of the genuinely large escape width, reproduced through the appropriate boundary conditions. 
The use of box boundary conditions (not shown) produces a structured distribution similar to the QRPA+D1S case. 
In $^{58}$Ni, the high-lying peak was observed at 34~MeV, while the calculated response function peaks around 30~MeV. 
In heavier nuclei theoretical models tend to predict a higher energy for the IS GDR than extracted experimentally~\cite{You2004b}. 
The reason for these discrepancies is likely a combination of factors, not restricted to the compression modulus, since discrepancies remain when the monopole compression mode is correctly described. 

At least two more structures, on which we shall remark next, are visible at lower energies: around 10~MeV and around $15-20$~MeV, depending on the isotope and the interaction. 

The lowest IS peak, typically close to 10~MeV, is the IS-LED mode (or surface mode), and we analyze it in Sec.~\ref{Sec:Two}. 
In $^{58}$Ni IS dipole strength has been detected in this energy region, as indicated in Fig.~\ref{Fig:ISGDR}(c). 
In $^{48}$Ni, this peak would lie within a broad continuum of proton-dominated transitions, as we infer from the CRPA+SLy4 results of Fig.~\ref{Fig:ISGDR}(a) and 
from the similar results analyzed in \cite{PPP2005}. 
The large width obtained with the CRPA+SLy4 model for the low-energy strength distribution of this nucleus reflects the proton escape width. 

The structure at $15-20$~MeV is no less interesting. 
Such an IS dipole peak, just below the IV GDR, has been detected in a variety of nuclei~\cite{Uch2004,You2004b}, including in $^{58}$Ni at 17.4~MeV. 
Its physical interpretation remains unclear~\cite{VWR2000,CVB2000,RRN2013}. 
A detailed study of this structure lies beyond the scope of the present work. 
In the following we will focus on the structures closer to and below 10~MeV, relevant to the physics of pygmy resonances. 

We may already observe that both models predict increased $E1$ strength in $^{68}$Ni and $^{78}$Ni 
around 10~MeV--cf. Fig.~\ref{Fig:sigma}(d),(e). 
Next we will identify transitions of potential interest which give rise to this and other effects.

 

\subsection{Systematics with neutron number\label{Sec:ResultsSystematic}} 

\begin{figure}
\includegraphics[width=0.4\textwidth]{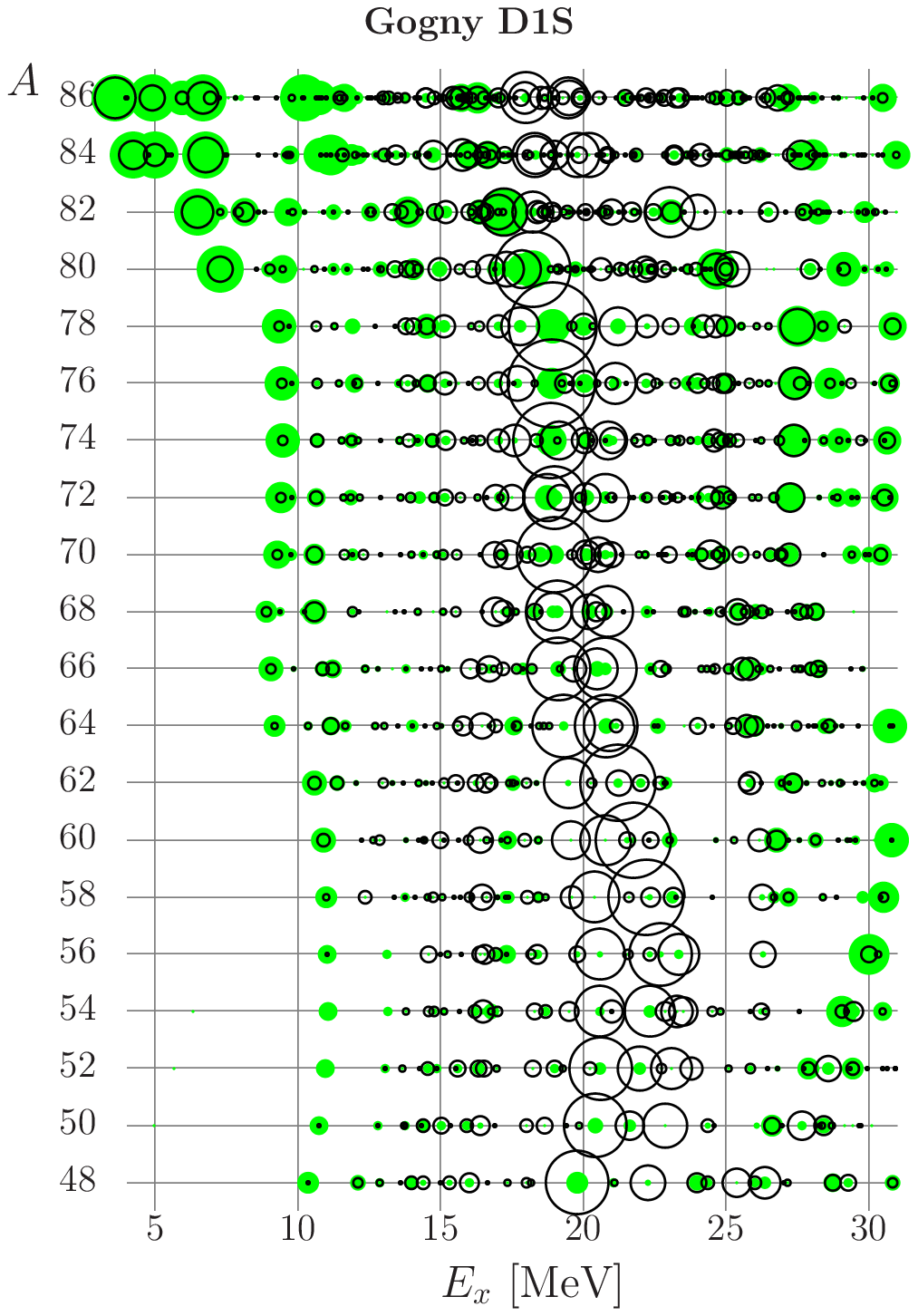}\\[2mm] %
\caption{(Color online) For the given isotopes and excitation energies, 
the IS and $E1$ transition strengths  
are indicated by green disks and black circles, respectively. 
The area of a disk or circle is propotional to the strength. 
 \label{Fig:Response}} 
\end{figure}

Fig.~\ref{Fig:Response} 
gives an overview of the response of all isotopes calculated with the QRPA+D1S method. 
The IS and E1 transition strengths  
are indicated by green disks and open black circles, respectively. 
The area of a disk or circle is propotional to the strength. 
The basic features of the spectrum and how they evolve with neutron number are clearly visible. 
The large circles in the energy region of $20\pm 3$~MeV correspond to the GDR, which  
appears fragmented into two or more structures. 
The lower-energy part of the GDR becomes strongly isoscalar in the more asymmetric isotopes, thus showing a mixed isospin character.  
The disks at high energy, above 25~MeV, correspond to the high-lying compression mode. 

The low-energy spectrum 
in the IS channel is dominated, in all isotopes, by at least one strong transition close to 10~MeV. 
We may interpret the lowest-energy peak, in all isotopes with $N=20-50$, 
as a mode of similar nature as the IS-LED studied in Refs.~\cite{PPR2011,PHP2012,PHP2014,Der2014}. 
We study its properties in Sec.~\ref{Sec:Two}.  
Even in $^{80,82}$Ni, the states visible at 9.5 and 9.7~MeV are candidates for such a mode. 

As neutrons are added to the $N=Z$ nucleus $^{56}$Ni, this transition appears to aquire more and more $E1$ strength. 
When we reach $^{62}$Ni we observe a splitting in the low-energy IS strength into two promiment transitions. 
Both appear to be of mixed isospin character, with the lower one remaining more strongly IS. 
The second structure presents an interesting candidate for an exotic mode generated by neutron excess. We will also analyse it in Sec.~\ref{Sec:Two}. 

In the proton-rich isotopes the IS-LED carries a very small amount of $E1$ strength. 
Based on our results with CRPA+SLy4 for $^{48}$Ni and those of Ref.~\cite{PPP2005}, we expect it to lie in the continuum and mix with proton transitions in most of these isotopes.   
Finally, beyond the shell closure of $N=50$ a large amount of IS and $E1$ strength is obtained at lower energies 
and it all is expected to lie in the continuum (see also the results in Refs.~\cite{HSZ1996,HSZ1998}) along with the IS-LED we identify in $^{80,82}$Ni.  
The nucleon separation energies in $^{80,82,84,86}$Ni, calculated within the D1S model, equal $2.05,1.88,1.84,0.71$~MeV, respectively.  

\subsection{How many modes?\label{Sec:Two}} 

We now examine the two types of low-energy states which we have identified as an isoscalar mode (IS-LED) excited in all isotopes, and a neutron mode in $^{62-76}$Ni. 

The coherent IS-LED mode of $^{56}$Ni has been analyzed in Ref.~\cite{PPR2011}, along with that of other $N=Z$ nuclei. 
The corresponding proton and neutron transition densities are very similar to each other and have a node, 
fitting well in the picture of the surface dipole mode of ordinary nuclei discussed in Ref.~\cite{Pap2015}.  
Let us now see to what extent such a picture persists in the other Ni isotopes. 

\begin{figure}
\includegraphics[width=0.36\textwidth]{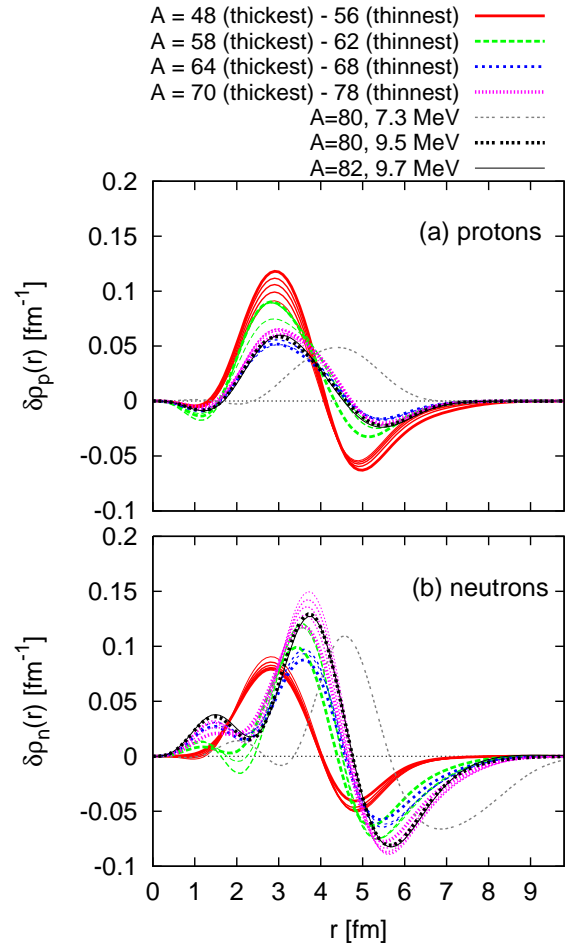} %
\caption{(Color online) 
(a) Proton and (b) neutron transition densities of the IS-LED mode in even Ni isotopes, 
calculated with the QRPA+D1S model. 
For comparison, the transition densities of the lowest dipole mode in $^{80}$Ni are also shown. 
\label{Fig:ISLED_All}} 
\end{figure}

In Fig.~\ref{Fig:ISLED_All} we show the proton and neutron transition densities for the IS-LED in all even isotopes with $A=48-82$, calculated within the QRPA+D1S model. 
For $A=48-78$ that is the lowest-lying IS state in Fig.~\ref{Fig:Response}. 
For $^{80,82}$Ni it is the IS state at 9.5 and 9.7 MeV, respectively. 
We notice that in all isotopes both transition densities are characterized by a node and they are in phase. 

The transition densities change very smoothly from $^{48}$Ni to $^{56}$Ni and from  $^{70}$Ni to $^{82}$Ni, 
and quite smoothly in between the two domains. 
In the neutron-deficient domain and including $^{56}$Ni  the similarities between the proton and neutron transition densities and in all isotopes are obvious. 
All these states can, therefore, be confirmed as of the usual IS-LED character.  

For $A\geq 60$ the neutron transition densities are stronger than the proton counterparts and somewhat dissimilar to them. 
Nevertheless, the surface protons contribute even in the very neutron-rich isotopes. 
The juxtaposition with the plotted lowest-lying state in $^{80}$Ni undescores this point.  
Therefore, to some extent we may also consider them as IS-LED states. 

\begin{figure}
\includegraphics[angle=-90,width=0.45\textwidth]{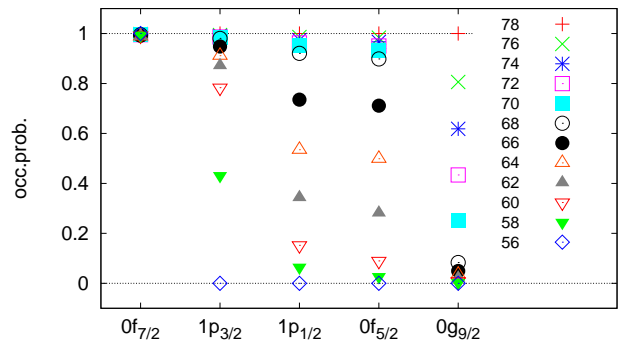} %
\caption{(Color online) 
Calculated occupation probabilities of the neutron $0f1p0g_{9/2}$ states in $^{56-78}$Ni. 
\label{Fig:OccNum}} 
\end{figure}

The observed discontinuities, with respect to the neutron number, especially visible in the neutron transition densities, can be attributed to shell structure, 
namely the filling of the neutron $pf$ shell and finally the $0g_{9/2}$ state. 
In reality, some of the discontinuities may be smoothened out by correlations beyond HFB and QRPA. 
The occupation probabilities of the above key orbitals, calculated within the HFB method, are shown in Fig.~\ref{Fig:OccNum}. 
From $^{60}$Ni to $^{68}$Ni the almost-degenerate (in this model) $1p_{1/2}$, $0f_{5/2}$ states fill up at a similar pace, while the $1p_{1/2}$ state is almost full and the $0g_{9/2}$ empty. 
Beyond $^{68}$Ni the strongest difference is the filling up of the $0g_{9/2}$ orbital. 

\begin{figure*}
\includegraphics[angle=-90,width=0.9\textwidth]{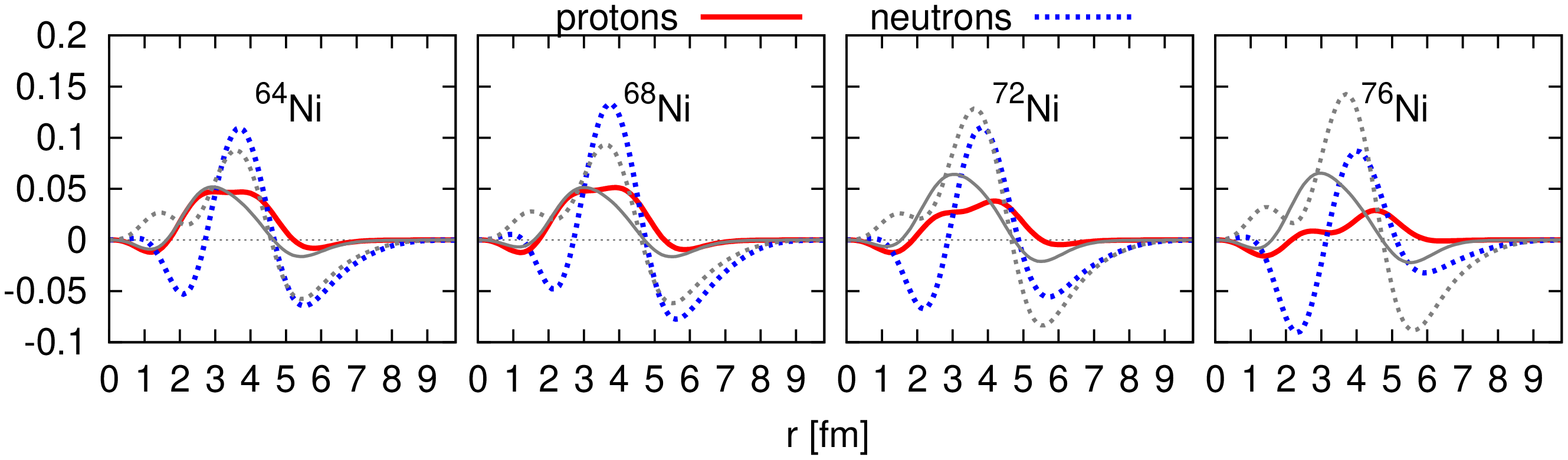} %
\caption{(Color online) 
Proton and neutron transition densities of the neutron-mode (or pygmy) candidate in $^{64,68,72,76}$Ni, 
calculated with the QRPA+D1S model. 
For comparison, the transition densities of the IS-LED mode (from Fig.~\ref{Fig:ISLED_All}) are also shown, plotted with thinner and paler lines. 
\label{Fig:PYGMY}} 
\end{figure*}

In Fig.~\ref{Fig:PYGMY} we show the transition densities of the neutron-mode candidates in $^{64,68,72,76}$Ni. 
For comparison, the IS-LED transition densities are also plotted with thinner lines. 
There are similarities between the two modes, in particular the presence of nodes and the non-negligible contribution of surface protons in most cases. 
We do notice, however, that the proton contribution to the neutron-mode candidate diminishes in the more neutron-rich isotopes -- see Fig.~\ref{Fig:PYGMY}(c),(d). 
Furthermore, this mode is predicted to decouple from the IS-LED and to develop only in neutron-rich Ni isotopes. 
For this reason we identify this as a separate mode, driven by neutron excess and assisted by shell effects. 
In particular, $^{62}$Ni  is the lightest isotope for which the $1p_{1/2}$ and $0f_{5/2}$ states are significantly occupied, activating new $2qp$ configurations. 
An analysis of the wavefunctions shows that the $0f_{7/2}\rightarrow 0g_{9/2}$ and the $1p0f\rightarrow 2s1d$ transitions are coupled into the IS-LED for the lighter isotopes, 
while for $^{62}$Ni and beyond they split from each other and $1p0f\rightarrow 2s1d$ attract and couple to the newly available excitations of $0g_{9/2}$ neutrons at somewhat higher energies. 

The lowest-lying states in $^{80-86}$Ni (below 8~MeV in Fig.~\ref{Fig:Response}) can be attributed to loosely-bound neutrons in spatially extended orbitals of the $sdg$ shell. 
These constitute yet another type of transition, similar to the soft modes of halo nuclei. 
The neutron transition densities extend well beyond 8~fm, while the proton transition densities remain localised and peak between 4 and 5~fm (surface-peaked). 
This is exemplified by the transition density of the 7.3~MeV state of $^{80}$Ni, shown in Fig.~\ref{Fig:PYGMY}. 
The strongly isoscalar and very weakly isovector transitions just above 10~MeV resemble the IS-LED, in that the proton transition densities have a node. 
We may therefore distinguish two groups of low-lying strongly isoscalar states beyond $^{80}$Ni, depending on how strongly isovector they are.  
Such broad structures might be observable, though not as individual transitions, because they are unbound. 

We now examine whether the IS-LED and the neutron mode in the lighter isotopes ($A\leq 78$) can be considered collective. 
The IS-LED is found strong in the IS channel, exhausting a few percentage points of the energy-weighted sum, 
and it energetically lies below the several $1\hbar\omega$ configurations contributing to its IS strength. 
We may, therefore, consider it as coherent--see also Ref.~\cite{PPR2011}. 
The neutron-mode candidate is quite strong in the IS channel around $^{68}$Ni and lies below most $1\hbar\omega$ configurations. 
Its $E1$ strength exceeds one single-neutron value, as we will see in Sec.~\ref{Sec:Sums} when we discuss Fig.~\ref{Fig:sumE1exp}. 
The above are signs of collectivity. 

\begin{figure}
\includegraphics[angle=-90,width=0.40\textwidth]{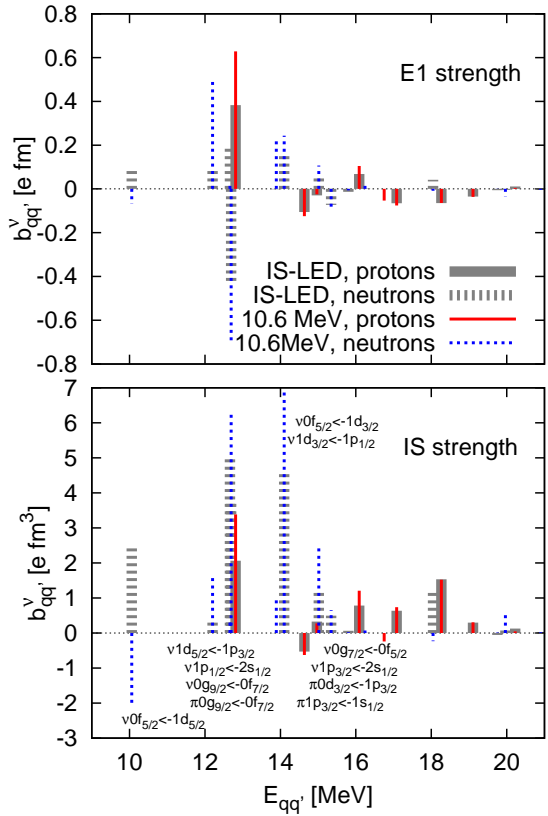} %
\caption{(Color online) 
The contributions of 2-quasiparticle configurations to the transition matrix element of the IS-LED and the $10.6$MeV mode in $^{68}$Ni, 
calculated in the QRPA+D1S model, 
(a) for the IV operator, Eq.~(\ref{eq:E1_IV_mod}), and 
(b) for the IS operator, Eq.~(\ref{eq:E1_IS_mod}).  
\label{Fig:XY}} 
\end{figure}

It is worth examining the wavefunctions of this type of mode. 
We will look at the matrix element terms $b_{qq'}^{\nu}$, Eq.~(\ref{Eq:bqqnu}),  
which determine the contribution of each configuration to the strength of the $10.6-$MeV mode in $^{68}$Ni. 
The above terms for the IV and IS operators are plotted in Fig.~\ref{Fig:XY}(a) and (b), respectively, as a function of the 2-quasiparticle energy. 
For comparison, the same quantities are shown for the IS-LED. 
We use the corrected operators, Eqs.~(\ref{eq:E1_IS_mod}),~(\ref{eq:E1_IV_mod}). 
Notice that if we considered the uncorrected IV operator in Fig.~\ref{Fig:XY}(a), 
only the proton contributions would appear in Fig.~\ref{Fig:XY}(a), multiplied by a factor $A/Z$, though the final result for the strength would be the same. 
Similarly, use of the uncorrected IS operator would alter the results in Fig.~\ref{Fig:XY}(b). 
We stress, therefore, that there is some ambiguity in analyses such as the following. 

The prominent proton and neutron contributions to the $E1$ strength at 12.8 and 12.7~MeV, respectively, in both modes come from the 
$f_{7/2}\rightarrow g_{9/2}$ 
configurations, which are not spatially extended. 
There are considerable destructive contributions from many other configurations.  
The neutron configurations at 12.2 and 13.9, 
namely 
$1p_{3/2}\rightarrow 1d_{5/2}$  
and $1s_{1/2}\rightarrow 1d_{3/2}$, 
are the ones which contribute more strongly to the neutron mode, than the IS-LED. 
They are low$-\ell$  configurations and therefore spacially extended. 
On the other hand, 
the $\nu 2s_{1/2}\rightarrow 1p_{1/2}$ configuration at 12.66~MeV contributes only to the IS-LED. 
 
In Fig.~\ref{Fig:XY}(b) we see that the IS contributions of the various $1\hbar\omega$ configurations are mostly coherent in both modes. 
We conclude from the above that both can be considered coherent IS modes, made up of fragments of many $1\hbar\omega$ configurations. 
In this respect our results are in agreement with those of Ref.~\cite{Roc2012}, obtained using Skyrme forces, 
at least as regards the sole mode analyzed there. 

Based on the above analysis, and on the much higher $E1$ strength of the second mode at 10.6~MeV, we may consider the latter as of a different kind, 
namely an excitation of neutrons on the nuclear surface. 
The presence of the 
$\nu f_{7/2}\rightarrow g_{9/2}$ 
configuration appears essential in generating coherence for this mode. 
Therefore the strength diminishes in $^{78}$Ni, where the neutron $0g_{9/2}$ state is filled. 

\begin{figure}
\includegraphics[angle=-90,width=0.50\textwidth]{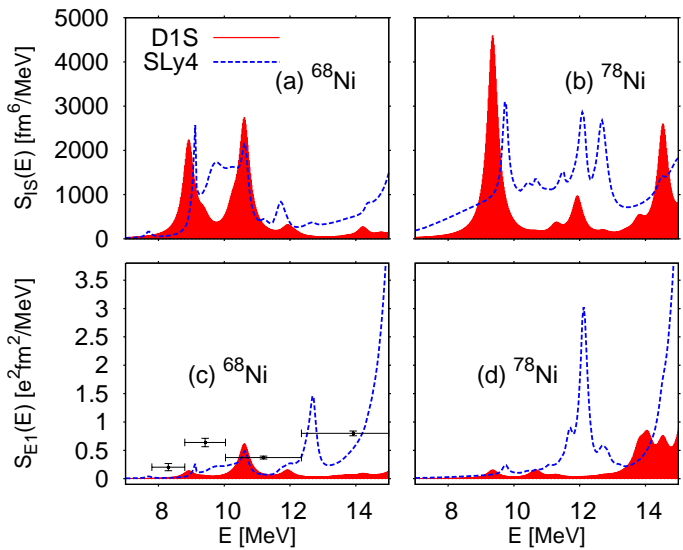} %
\caption{(Color online) 
The IS and $E1$ strength distributions, for $^{68}$Ni and $^{78}$Ni, calculated with the QRPA+D1S and CRPA+SLy4. 
The response functions are smoothed with a width of 0.4~MeV. 
The data in (c) are from Ref.~\cite{Ros2013,AuRPC}. 
\label{Fig:ResponseNi6878}} 
\end{figure}

In Fig.~\ref{Fig:ResponseNi6878} we show the IS and $E1$ strength distribution of $^{68}$Ni and $^{78}$Ni in this energy region, 
calculated with both models. 
We notice that the bimodal structure of the response of $^{68}$Ni 
is present also in our results with the CRPA+SLy4 model. 
Noteworthy are the significant contributions from the non-resonant continuum in both channels in the region up to around 11~MeV. 
The effect is predicted to be especially important in $^{78}$Ni. 
%
Such continuum transitions, first studied in the 1990's (see, e.g., \cite{HSZ1996,HSZ1998}), 
are likely critical for describing correctly the suprathreshold strength beyond $^{68}$Ni. 
In fact, in the case of $^{68}$Ni, if both resonances, namely the IS-LED and the neutron mode, are bound transitions,  
the QRPA+D1S cannot explain the observed strength. 
The same can be said about other configuration-space calculations, employing the SLy4 or other Skyrme functionals~\cite{TeE2006}. 
In general, the energies of the unbound single-particle states, and therefore the non-collective single-particle spectrum in the continuum, converge very slowly with the 
harmonic-oscillator basis size. 
In actual calculations they can never be considered converged.  
If the particle threshold is much lower than the GDR, this shortcoming becomes relevant for pygmy strength. 


It is noteworthy that the CRPA+SLy4 model predicts increased $E1$ strength also around $12-13$ MeV, compared to the QRPA+D1S model, in the form of additional peaks. 
Peaks in this energy region are obtained for all four isotopes studied with the CRPA-SLy4 model, as is evident in Fig.~\ref{Fig:sigma}.  
As the most possible reason for the additional CRPA strength in that energy region 
we find the overall energetic displacement of one GDR strength distribution with respect to the other. 
The proton and neutron transition densities 
$r^2\delta\tilde{\rho}^{\mathrm{IV}}_{p,n}(r)$ of the CRPA peaks in question, for $N\geq Z$, typically have prominent peaks at a value of $r$ somewhat larger than the mean-square radius 
and they are out of phase at that $r-$value.%
\footnote{In $^{48}$Ni (and to a large extent in $^{78}$Ni) the surface protons (neutrons) play a dominant role.}  
Then the peaks could be considered as fragments of the GDR. 
They are, however, embedded in a continuum of transitions dominated by surface neutrons. 
We note that the existing data for $^{58}$Ni (Fig.~\ref{Fig:sigma}(b)) and $^{60}$Ni (not shown) do not present a prominent structure on the lower-energy tail of the GDR. 
We obtained similar results for the GDR and for these peaks using the SkM$^{\ast}$ functional instead of the SLy4.  
A similar fragmentation pattern is observed even with the QRPA+D1S model at somewhat higher energies.  


\subsection{$E1$ transition strength and polarizability\label{Sec:Sums}} 

\begin{figure}
\includegraphics[angle=-90,width=0.48\textwidth]{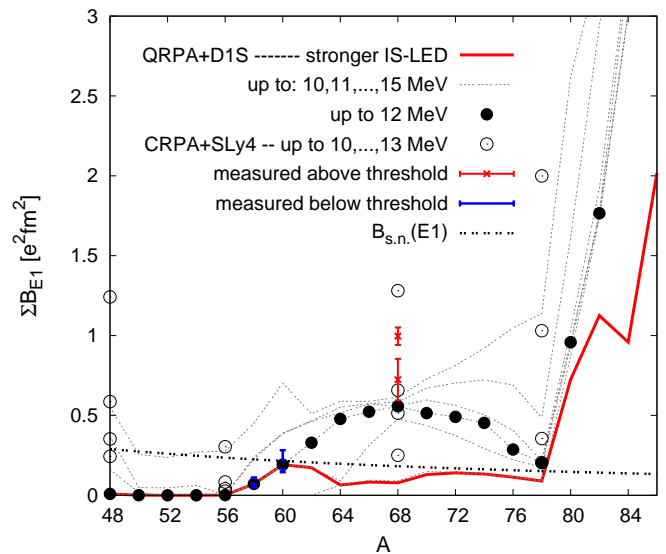}\\[2mm] %
\caption{(Color online) Sum of $E1$ strength for the indicated energy cuttoffs as a function of mass number. 
For $^{58,60}$Ni, data shown are from Refs.~\cite{Bau2000,Sch2013}. 
The data points for $A=68$ are: an estimate assuming a single peak with the reported mean energy and energy-weighted strength~\cite{Ros2013}  
(lower point) and the actual sum of strength measured between 7.8 and 10~MeV~\cite{AuRPC} (upper point). 
An earlier measurement of the energy-weighted summed strength~\cite{Wie2009,Car2010} gives the estimate $1.1\pm 0.3e^2$fm$^2$.  
The single-neutron unit, $B_{\mathrm{s.n}}(E1)$, is given by Eq.~(\ref{Eq:Bsn}). 
\label{Fig:sumE1exp}} 
\end{figure}

The $E1$ strength of the IS-LED and the summed $E1$ strength for different energetic cutoffs as a function of mass number is shown in Fig.~\ref{Fig:sumE1exp}.   
First we discuss our results with the QRPA+D1S model. 
We observe that the $E1$ strength of the IS-LED rises up to $A=60$ and then drops again, as a result of the splitting already seen above. 
An interesting result is that the summed strength up to 12~MeV or similar value 
does not rise monotonically, but it peaks close to $^{68}$Ni. 
Similarly, it has been observed that the low-energy $E1$ strength does not rise monotonically between $^{40}$Ca and $^{48}$Ca~\cite{Har2004}.  
Clearly, the neutron excess is not the only mechanism determining the $E1$ strength. 
In the context of our calculations, the explanation for the above behavior must lie in the shell structure. 
As we already mentioned, 
between $^{68}$Ni and $^{78}$Ni the $0g_{9/2}$ state fills up, blocking transitions from the $0f_{7/2}$ state. 
Although transitions from the $0g_{9/2}$ are available for $N>40$, we found that they contribute little strength in this region. 
The above anticorrelation between the strength and the neutron number in neutron-rich Ni isotopes was found also in Ref.~\cite{LCM2007} and attributed to the filling of the $0g_{9/2}$ orbital. 

The strength of the neutron mode in $^{68}$Ni, analyzed in Sec.~\ref{Sec:Two}, exceeds the single-particle estimate. 
The summed $E1$ strength in this region was found large in Sn isotopes too~\cite{PHP2014}, but in those cases it is generated by individual weak transitions. 

Given our past results and how they compare with data~\cite{PHP2012,Der2014,PHP2014}, 
we expect the summed strength up to 12~MeV to correspond to a summed experimental strength up to, roughly, 9 MeV. 
For the stable isotopes this would mean an excellent agreement with the data shown in Fig.~\ref{Fig:sumE1exp}. 
In $^{68}$Ni, pygmy strength was found in the continuum up to 12~MeV. 
Notice that, again based on the known trends of the model, we expect the first IS peak to be bound, i.e., below the threshold energy of 7.8~MeV.  
(For $^{72}$Ni and heavier isotopes, for which the neutron threshold is below 7~MeV, practically all strength is likely in the continuum.)  
It follows that our results are marginally compatible with the data in the case of $^{68}$Ni. 

Now we turn to the $E1$ strength obtained with the CRPA+SLy4 model, also shown in Fig.~\ref{Fig:sumE1exp}, for various summation cutoffs. 
We notice that the predictions of this model agree with the data for $^{68}$Ni better than those of the QRPA+D1S model. 
A much more dramatic difference between the two models is seen in $^{78}$Ni (cf. Fig.~\ref{Fig:ResponseNi6878}). 
Threshold transitions described efficiently by the CRPA model likely become increasingly important beyond $^{68}$Ni. 

\begin{figure}
\includegraphics[angle=-90,width=0.45\textwidth]{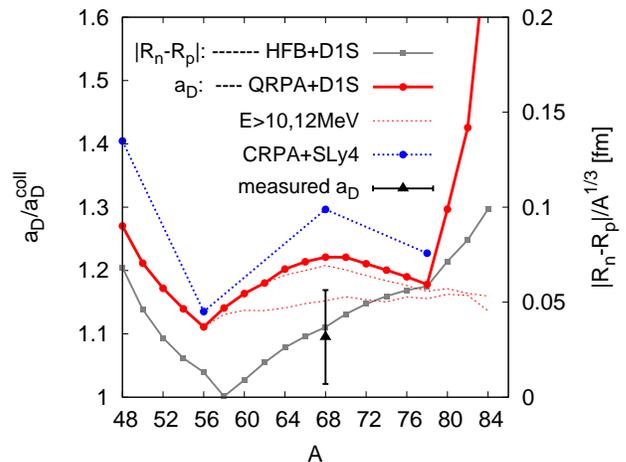}\\[2mm] %
\caption{(Color online) The electric dipole polarizability $a_D$ of Ni isotopes (left axis) calculated with both models as indicated and 
the difference of the proton and neutron root-mean-square radii (right axis) calculated with the HFB+D1S model. 
The polarizability is given relative to the estimate $a_D^{\mathrm{coll}}$ for a single collective mode. 
The radii difference is divided by $A^{1/3}$.  
The measurement of $a_D$ in $^{68}$Ni is from Ref.~\cite{Ros2013}. 
\label{Fig:polariz}} 
\end{figure}

Finally we present results for the dipole polarizability $a_D$ of the Ni isotopes. 
In Fig.~\ref{Fig:polariz} we plot $a_D$ relative to the reference value $a_D^{\mathrm{coll}}$ defined in Eq.~(\ref{Eq:aDcoll}), to eliminate trivial dependencies on the mass number. 
The polarizability as a function of neutron number varies in a smooth manner between shell closures, at which points, namely $^{56}$Ni and $^{78}$Ni, it presents kinks. 
Obviously this quantity is influenced by shell structure. 
Beyond $^{78}$Ni it rises rapidly. 
This behavior can be attributed to the strong low-lying transitions observed in Fig.~\ref{Fig:Response}.  
Indeed, if we calculate $a_D$ without taking into account the lower-lying states, the function $a_D(A)/a_D^{\mathrm{coll}}(A)$ becomes smoother and follows closely the linear trend, as evident in Fig.~\ref{Fig:polariz}. 
 
We notice that the two models overestimate the polarizability of $^{68}$Ni, even though the functionals are asy-soft and underestimate the strength of the observed pygmy state. 
We elaborate on this contradiction at the end of Sec.~\ref{Sec:Asy}. 

In Fig.~\ref{Fig:polariz} the neutron or proton-skin thickness, defined via the root-mean-square radii 
\[ 
| R_n-R_p | \equiv 
\left| 
\sqrt{\langle r^2 \rangle_n} - 
\sqrt{\langle r^2 \rangle_p}  
\right| 
\] 
is also shown as a function of $A$. 
In particular, we plot the quantity 
$ | R_n-R_p | /A^{1/3} $, which    
represents a skin thickness relative to the nuclear radius. 
It is calculated in the HFB model with the Gogny D1S interaction.  
The value obtained for the neutron-skin thickness in $^{68}$Ni is 0.15~fm. 
The proton and neutron radii are found almost equal in $^{58}$Ni, while for $A\leq 56$, i.e., $Z\leq N$,  a proton skin develops. 
%
We notice that the (proton or neutron) skin thickness increases monotonically with the proton-neutron asymmetry, except around $N=Z$. 
The effect of neutron deficiency is stronger when the continuum is propertly treated (CRPA+SLy4). 
Importantly, the fact that the relative polarizability decreases beyond $^{70}$Ni shows that it does not correlate with the relative neutron-skin thickness, but it is strongly influenced by shell effects. 
A possible correlation emerges beyond the $N=82$ shell closure -- see also Ref.~\cite{ENI2013a}. 
We conclude from the above that the relation between the proton-neutron asymmetry and the polarizability, along an isotopic chain, is not completely trivial.

\section{Critical discussion in view of other existing calculations\label{Sec:Discuss}}  

\subsection{Ca, Sn, isospin splitting, and the importance of shell effects\label{Sec:Shell}} 

We now compare results for the three isotopic chains, Ca~\cite{PHP2012,Der2014}, Ni (this work), and Sn~\cite{PHP2014}. 

We begin with the following general observation, which will help us interpret the isospin structure of pygmy strength and especially the so-called isospin splitting in heavy stable nuclei. 
By the latter term we mean the phenomenon whereby $E1$ transitions are observed over a range of excitation energies below particle threshold, 
of which only the lowest-lying respond to an IS field~\cite{SAZ2013}.  

First we observe, in the compilation of Ref.~\cite{Pap2015}, that the energy of the isoscalar segment, typically $6-7$~MeV, varies weakly with mass number in comparison, e.g., to the shell energy $1\hbar\omega =41A^{-1/3}$MeV. 
Pygmy dipole strength, according to non-relativistic models, is expected at around $1\hbar\omega$~\cite{Bar2013} and can be attributed to fragments of $1\hbar\omega$ transitions, rather than a collective mechanism~\cite{OHD1998,KrS2009,Roc2012,ReN2013}.  
In medium-mass stable nuclei, such as stable Ca and Ni isotopes, the IS and pygmy ($1\hbar\omega$) transitions generally lie far apart, 
because of the above $A$ dependence. 
In heavier stable nuclei, at least the lower-energy tail of the pygmy strength distribution can approach and overlap with the IS state. 

Indeed, a noteworthy difference observed in the experimental data on stable nuclei is that just a few weak IV transitions are seen on either side of the IS-LED in $^{48}$Ca~\cite{Der2014}, while 
the IS-LED candidate lies on the lower end of the, less weak, $E1$ spectrum in stable Sn isotopes~\cite{PHP2012,End2010} and other heavy nuclei~\cite{SAZ2013}.  
A plausible explanation for the so-called isospin splitting of pygmy strength in heavy stable nuclei is then the energetic proximity of two resonance-like structures: the IS-LED and the tail of the higher-lying pygmy mode (whether collective or not). 

We conclude from the above that the isospin character of the dipole strength distribution at low energies depends on the mass region. 

Regarding the isospin structure of the $E1$ states in stable Ni isotopes, below threshold, 
we expect that the IS and (weak) IV strength distributions will mostly coexist without a clear energetic splitting. 
Our expectation is based on the extended IS strength distribution detected in $^{58}$Ni~\cite{Poe1992}. 
The situation would resemble that in $^{48}$Ca, but with a fragmented IS strength distribution. 
We cannot conclude on the matter based on our calculations, since QRPA predicts the IS-LED always to be the lowest-lying dipole mode (see Ref.~\cite{Der2014} regarding $^{48}$Ca) and does not include fragmentation mechanisms like phonon coupling~\cite{LRT2013,Pon2014,Kna2014,Ach2015}. 

Next we turn to a different observation. 
In the Ca and Sn isotopic chains we found that the $E1$ strength of the IS-LED state does not increase monotonically as we add only a few neutrons to the $N=Z$ isotope, 
but it reaches a minimum for $N$ slightly larger than $Z$. 
Since the $E1$ strength of this IS mode is a result of local dissimilarities between the proton and neutron transition densities~\cite{PPR2011}, 
reflecting, in turn, dissimilarities in the ground-state densities, it appears easy to explain its behavior as an effect of the Coulomb interaction: 
For $N=Z$ the proton density distribution should be slightly more extended than the neutron density distribution, leading to this local dissimilarity. 
Addition of a few neutrons reverses the situation, resulting in minimal $E1$ transition strength in between, at some small value of $N-Z$. 

This interpretation, which relies on a semi-classical image of density profiles, 
breaks down in the Ni chain, where minimal $E1$ strength is obtained for $N$ slightly {\em smaller} than $Z$, signifying a competition between Coulomb and asymmetry effects, on the one hand, and shell effects on the other. 
Indeed, the Ni shell structure tends to produce the opposite effect to asymmetry: 
Addition of neutrons to $^{56}$Ni means filling up lower-${\ell}$ states, which are extended to the nuclear surface, immediately initiating what resembles a neutron-skin effect~\cite{WCV2014}. 
Removal of neutrons from the $0f_{7/2}$ state, on the other hand, creates some neutron deficiency towards the interior, rather than the exterior, thus delaying the onset of  the proton-skin effect 
in the dipole response. 
The above asymmetry between neutron excess and neutron deficiency in the case of the Ni isotopic chain is obvious in the transition densities of the IS-LED, shown in Fig.~\ref{Fig:ISLED_All}. 
 
Finally, the splitting of the low-energy part of the response, from about $^{62}$Ni to the shell closure, 
and the corresponding unsmooth behavior of the IS-LED properties, 
 constitutes a departure from what we observed, with the same model, 
in the Ca and Sn isotopic chains. 
In Ca, the filling of the $\nu 0f_{7/2}$ subshell from $^{40}$Ca to $^{48}$Ca results in only a moderate and smooth change in the properties of the IS-LED. 
All involved nuclei are stable. 
In Sn, the filling of the $\nu 2s-1d-0g_{7/2}-0h_{11/2}$ shell gives the same result, even though many of the nuclei are unstable. 

All the above observations underscore the relevance of shell structure in determining the $E1$ strength distribution at low energies. 
Furthermore, shell effects such as the above have been found to influence the degree of correlation between the neutron-skin thickness 
and the amount of low-energy $E1$ strength~\cite{INY2011,INY2013,ENI2013a,ENI2014}. 

\subsection{Pygmy states, polarizability, and asy-stiffness\label{Sec:Asy}}

Next, we comment on the possible role of the symmetry energy and its slope $L$ (asy-stiffness) in determining the low-energy $E1$ strength distribution, 
and in particular   
how our results fit into the ongoing debate.  
A more profound discussion would require further quantitative analyses and, as we argue eventually, more accurate and complete data on $^{68}$Ni. 


Our study of the Ca isotopic chain~\cite{PHP2012,Der2014}  showed than no skin mode develops in $^{48}$Ca at low energies (we cannot exclude one above threshold). 
The Gogny interaction, being asy-soft, predicts correctly that the IS-LED mode survives as such up to $^{48}$Ca. 
Now we have found that the QRPA+D1S model underestimates the strength of the pygmy resonance, as measured around 10~MeV in $^{68}$Ni, perhaps by a large amount.  
This could suggest that the functional is too soft -- hardly a novel conclusion~\cite{Car2010}. 
The same will be true for the SLy4 functional (also asy-soft), if the IS-LED and the neutron mode are bound. 

The effect of stiffness on the low-energy strength in $^{208}$Pb, $^{132}$Sn, as well as $^{68}$Ni, can be observed in the results of Ref.~\cite{Roc2012}: 
In view of the measured strengths, the SkI3 functional ($L=100.5$MeV) is too stiff, while the SLy4 ($L=46.0$MeV) and SGII ($L=37.7$MeV) functionals, 
and even more so the D1S ($L=22.4$MeV), are too soft, in line with the conclusions of Ref.~\cite{Car2010}.  
We note also that relativistic functionals, which are typically more asy-stiff~\cite{Pie2002,WVR2009}, 
predict strong neutron-skin modes, in the form of neutron-skin oscillations, for moderately asymmetric nuclei, 
thus overestimating their low-lying strength. 

We now point out that the softness effect is difficult to disentangle from shell effects. 
Extracting the asy-softness from data on a single nucleus, like $^{68}$Ni, would be questionable, 
even though the correlations within a given model are found strong~\cite{ReN2010}. 
Rather, a different strategy to connect data with the symmetry energy and softness, and subsequent improvement of functionals, suggests itself, akin to the spirit of Ref.~\cite{Car2010}: 
The functional must be soft enough to reproduce the small $E1$ strength observed in the stable Ca and Sn isotopes below $8-9$~MeV, as well as $^{208}$Pb. 
At the same time, it should generate the correct amount of $E1$ strength in $^{68}$Ni above threshold. 
Accurate data on $^{68}$Ni are highly desirable. 
Furthermore, data below threshold would clarify the origin of the strength observed above threshold, with the help of our present results. 
 

An important related observation is that, although both models used here are rather soft, they overestimate the polarizability of $^{68}$Ni. 
The same does not hold for other nuclei. 
In Table~\ref{Table} we compare results obtained with the two models and measured values of $a_D$ in 
$^{68}$Ni~\cite{Ros2013},  
$^{120}$Sn~\cite{Has2015X}, 
and $^{208}$Pb~\cite{Tam2011}: 
\begin{table}  
\begin{center}  
\begin{tabular}{|l|ccc|} 
\hline  
   & Exp      & D1S    &  SLy4 \\ 
\hline 
$^{208}$Pb & 20.1(6)  & 18.78  & 20.075 \\
$^{120}$Sn & 8.93(36) & 8.45   &  9.40  \\
$^{68}$Ni 
& 3.40(23)$^{\ref{f:dip}}$ 
& 3.79   &  4.03  \\     
\hline 
\end{tabular} 
\end{center} 
\caption{\label{Table}Dipole polarizability of three nuclei, in units of fm$^3$,  calculated within the QRPA-D1S and CRPA-SLy4 models, and measured values (``Exp") 
from Refs.~\cite{Tam2011,Ros2013,Has2015X}. 
}  
\end{table} 
The QRPA+D1S model underpredicts the polarizability of $^{120}$Sn and $^{208}$Pb by 5\% and 7\%, respectively, while  
the CRPA+SLy4 model describes accurately the polarizability of $^{208}$Pb and overpredicts that of $^{120}$Sn by 5\%. 
Both models overestimate the polarizability of $^{68}$Ni by 10\% or more. 
If we trust the model systematics, namely that such soft models cannot overestimate the dipole polarizability by such a factor, then 
the discrepancy represents a large amount of $E1$ strength potentially missing 
from the data.\footnote{\label{f:dip}It is reported~\cite{AuRPC} that a revised measured value of 3.88(31)~fm$^3$ is obtained by extrapolating a Lorenzian fit to the data~\cite{Ros2013} beyond the measured energy window. 
That could lift some of the discrepancy observed here. 
} 


We iterate that, besides symmetry and isovector properties of the nuclear energy-density functional, 
additional factors determine the strength distribution, which include shell effects and phonon coupling, 
as well as the nucleon effective mass in the case of mean-field functionals and the convergence of the calculations just above particle threshold. 
We conclude, in any case, that more accurate and complete data are needed for $^{68}$Ni, 
than reported in Ref.~\cite{Ros2013}, before the polarizability of this nucleus can be used 
to calibrate theoretical models or constrain the equation of state with confidence.


\section{Summary and conclusions\label{Sec:Summary}} 

We have calculated the dipole response of even Ni isotopes using the QRPA+D1S Gogny model and the CRPA+SLy4 Skyrme model, 
which predict similar nuclear-matter properties of relevance (symmetry energy and softness, nucleon effective mass). 
We have analyzed thoroughly the dipole transitions at low energy. 
Our main conclusions can be summarized as follows: 
\begin{itemize} 
\item 
The two models generally agree in their predictions regarding the various transitions below 11~MeV, except for additional continuum strength in the case of CRPA. 
Above 11~MeV and in the region of the GDR 
there are discrepancies in the IV channel, attributable to the different TRK enhancement factor. 
\item 
The summed $E1$ strength of stable isotopes at low energies is well reproduced. 
\item 
In all isotopes, a strong IS low-energy dipole mode (IS-LED) is obtained around 10~MeV with the QRPA+D1S model. 
Given the systematics of our model, we expect it to actually occur around 7~MeV. 
In the stable and moderately neutron-rich isotopes it is expected to be a bound transition. 
\item 
We showed that shell structure plays an important role in determining the low-energy $E1$ strength distribution. 
Furthermore, we conjectured that the isospin structure of low-energy strength, observed in stable nuclei, is determined by the difference between the IS-LED energy and the $1\hbar\omega$ value.  
\item 
In $^{68}$Ni and neighboring isotopes, besides the IS-LED, one more coherent IS transition is found, lying a couple of MeV above the IS-LED. 
Complete data are needed to determine whether there is a bimodal structure below threshold, or whether we can associate this second peak with 
the one detected above threshold~\cite{Wie2009,Ros2013}. 
\item 
Beyond the shell closure, $^{78}$Ni, the neutron-rich isotopes are found highly polarizable, due to valence-neutron transitions. 
\item 
We showed that the relation between the polarizability and the skin thickness along an isotopic chain is not straightforward, 
because they are influenced by shell effects in different ways or degrees. 
\item
The QRPA+D1S model underestimates the threshold $E1$ strength in $^{68}$Ni. 
Furthermore, if both modes identified above (IS-LED and neutron mode) are bound, the model fails to explain the peak observed above threshold. 
One reason for these problems must be the asy-softness of the functional, or the residual interaction in the IV channel, pushing too much strength into the GDR. 
Other reasons may include the nucleon effective mass, effects beyond QRPA, especially phonon coupling~\cite{LRT2013,Pon2014,Kna2014,Ach2015}, and the continuum treatment, the latter of which we examined. 
\item 
We found that a converged treatment of continuum threshold states, as in the CRPA model, leads to additional threshold strength in neutron-rich nuclei. 
In $^{78}$Ni its magnitude could be significant. 
\item 
Although the present functionals are soft, mostly underestimate the dipole polarizability of stable nuclei 
and underestimate the $E1$ strength in $^{68}$Ni, they overestimate the polarizability of $^{68}$Ni. 
This leads us to infer that strength is missing from the $^{68}$Ni data reported in \cite{Ros2013}.  
%
\item 
We conjectured, pending further systematic studies, 
that accurate and complete data on $^{68}$Ni not only could clarify the above issues, but provide a complementary constraint to the asymmetry stiffness from below, 
while data on stable nuclei could constrain it from above.  
\end{itemize} 
In the future, it will be very interesting to confirm experimentally the dramatic role of shell closure revealed in this and other theoretical studies. 
This would require data along isotopic chains, in particular nuclei on either side of a shell closure, and in the same energy range, especially in relation to the particle emission threshold. 

In short, the present study of the Ni chain, with its particular shell structure and many isotopes, has helped us identify new phenomena and factors 
which determine the low-energy dipole strength. 
New data above and below particle-emmision threshold should enable us to test our conclusions and provide important feedback for nuclear structure theory.

\begin{acknowledgments}
We thank T.Aumann and D.Rossi for communicating their data on $^{68}$Ni. 
This work was supported 
by the Rare Isotope Science Project 
of the Institute for Basic Science 
funded by the Ministry of Science, ICT and Future Planning 
and the National Research Foundation of Korea (2013M7A1A1075766), 
by the National Superconducting Cyclotron Laboratory, 
by the U.S. Department of Energy, Office of Science, Office of Advanced Scientific Computing Research, through the NUCLEI SciDAC-3 Collaboration, 
by the Deutsche Forschungsgemeinschaft through contract SFB 634, 
by the Helmholtz International Center for FAIR (HIC for FAIR), 
and 
by the BMBF through contract 06DA7074I. 
Computing resources were provided by the Michigan State University High Performance Center (HPCC) / Institute for Cyber-Enabled Research. 
\end{acknowledgments}



%

\end{document}